# Simplified Bond Hyperpolarizability Model of Second Harmonic Generation, Group Theory and Neumann's Principle

Adalberto Alejo-Molina
Center for surface and nanoanalytics, Johannes Kepler University, Altenbergerstr. 69, 4040 Linz, Austria
Instituto Nacional de Astrofísica Óptica y Electrónica, Luis Enrique Erro No. 1, 72840 Tonantzintla, Puebla, Mexico
E-mail: adalberto_am@hotmail.com

Hendradi Hardhienata
Center for surface and nanoanalytics, Johannes Kepler University, Altenbergerstr. 69, 4040 Linz, Austria
E-mail: hendradi_h@yahoo.com

Kurt Hingerl
Center for surface and nanoanalytics, Johannes Kepler University, Altenbergerstr. 69, 4040 Linz, Austria
E-mail: kurt.hingerl@jku.at

**Abstract.** We discuss the susceptibility third-rank tensor for second harmonic and sum-frequency generation, associated with low index surfaces of silicon (Si(001), Si(011) and Si(111)) from two different approaches: the Simplified Bond-Hyperpolarizablility Model (SBHM) and Group Theory (GT). We show that SBHM agrees very well with the experimental results for simple surfaces because the definitions of the bond vectors implicitly include the geometry of the crystal and therefore the symmetry group. However, for more complex surfaces it is shown that one can derive from GT the SBHM tensor, if Kleinman symmetry is allowed.

## I. Introduction

Nonlinear Optics (NLO) is a field of optics which experienced rapid progress not only in theory but also experimentally with many important direct applications in telecommunication, sensing, material characterization, and various other fields. In NLO, the optical properties of a material can be modified by light itself, given that the incoming light has a high intensity such as a laser. The nonlinearity arises from the material response to the incoming light field, which is not only proportional to the electrical field, but also to its higher powers (e.g. the square of the electrical field or superior powers of the amplitude). The nonlinear response that produces light with twice the incoming light frequency is called second harmonic generation (SHG). It follows therefore that the amplitude of the second harmonic frequency generated by a sample is proportional to the square of the incoming laser light field. Another consequence of SHG is that the refractive index depends also on the intensity of the incoming light which is contrary to the linear case.

Second order nonlinear radiation can only occur in non-centrosymmetric crystals because these crystals can have uneven powers in its potential, thereby even powers in the fields/forces. Common materials such as conventional liquids, gases, amorphous solids (glasses) and many crystals exhibit inversion symmetry which requires that all the elements in the third rank tensor $\chi^{(2)}$ are zero. Therefore, the main contribution to second harmonic generation due to the electronic dipoles, cannot occur in such system. However, as it is well known, there are several mechanism which can produce small second harmonic signals such as retardation, spatial-dispersion, and magnetic effects, in addition to the anharmonic restoring force acting on the bond charge [1]. Other works also included possible SHG sources due to bulk quadrupole notably the work of J. Kwon and co-workers [2] using a fourth rank tensor. However since these sources are not directly linked to the crystal symmetry they are not treated in our contribution discussing the third rank tensor structure.

In the surface or interface however, the inversion symmetry is broken and is represented by a nonzero $\chi^{(2)}$, thus second order nonlinear processes such as SHG and SFG are allowed for example at solid/air, liquid/liquid, liquid/air, semiconductors, polymer and biological films [3-6]. Hence, the main SHG contribution in centrosymmetric crystals only comes from a few layers near the surface, which feel the disturbance for the deviation of periodicity. However, even if it is only considered that the SH signal is generated from the surface/interface, the mathematical model is still complicated and requires a third rank tensor with up to 27 complex components, making it very difficult to interpret experimental data and develop an easy and accurate phenomenological model. In the past years, several models to describe SHG were proposed [7-11]. In particular, the simplified bond-hyperpolarizability model (SBHM) applied to Si crystal surfaces is very interesting because the physical mechanism of SHG is stated clearly in classical terms and only requires a few free parameters to model the data obtained by experiment [9, 10].

In this paper we investigate how group theory (GT) is commonly used to simplify the third rank susceptibility tensor and explain why SBHM is mathematically equivalent to this approach. Therefore, the systematic of this paper is as follows: In section 2, we discuss the assumptions and the main results of the third rank tensor using SBHM. Section 3 focuses on a brief explanation of group theory and its applications to crystallography as well as how it is related to SBHM. Finally, the conclusions of this work are presented.

**II. Main Results of the Simplified Bond-Hyperpolarizability Model**

SBHM was first proposed by D. Aspnes and co-workers in 2002 [9, 10]. The essence of this model is to assume that the nonlinear polarization source produces an anharmonic dipole oscillation along the atomic bonds, assigning different values for the polarizabilities and hyperpolarizabilities according to their direction (i.e. pointing towards the vacuum side or bulk side). In addition, SBHM in its original formulation [9, 10] also assumes that the electron moves only along the bond. Furthermore, it is assumed that the bond orientations in the surface have no molecular reconstruction. This simple model has been proved to describe very well the azimuthal SHG intensity dependence for (111) and (001) Si-dielectric interfaces [9-11]. At the present there are more complicated models that for example take into account of the transversal motion of the charges along the bonds and the charge distribution with cylindrical shape but their contribution to the final polarizability vanish for SHG [7, 8]. Further extensions cope with bulk quadrupole effects in the spirit of SBHM, which has been developed by Kwon *et al.* [2] to produce some previously unexplainable results for SHG from the Si (001) crystal.

In SBHM the polarization up to the second order is calculated by [9]:

$$\vec{P} = \frac{1}{V}\sum_j \left[\alpha_{1j}\mathbf{R}^{(z)}(\phi)\cdot\hat{b}_j\right]\cdot\vec{E} + \frac{1}{V}\sum_j \left[\alpha_{2j}\left(\mathbf{R}^{(z)}(\phi)\cdot\hat{b}_j\right)\otimes\left(\mathbf{R}^{(z)}(\phi)\cdot\hat{b}_j\right)\otimes\left(\mathbf{R}^{(z)}(\phi)\cdot\hat{b}_j\right)\right]\cdot\cdot\vec{E}\otimes\vec{E} \quad (1)$$
$$= \chi^{(1)}\vec{E} + \vec{\chi}^{(2)}\vec{E}\vec{E}$$

therefore, the susceptibility third rank tensor, is calculated by the formula

$$\vec{\chi}^{(2)} = \frac{1}{V}\sum_j \alpha_{2j}\left(\mathbf{R}^{(z)}(\phi)\cdot\hat{b}_j\right)\otimes\left(\mathbf{R}^{(z)}(\phi)\cdot\hat{b}_j\right)\otimes\left(\mathbf{R}^{(z)}(\phi)\cdot\hat{b}_j\right) \quad (2)$$

where $V$ is the volume, $\alpha_{1j}$ are the linear polarizabilities, $\alpha_{2j}$ are the hyperpolarizabilities and $\hat{b}_j$ are the unit vectors in the direction of the atomic bonds. Note that the summation is going over all the bonds in the conventional cell. To simulate the experiment where the sample is rotated along the *z*-axis, we introduce the rotation matrix $\mathbf{R}^{(z)}(\phi)$.

In this work, we are going to represent a general third rank tensor as a 9 × 3 matrix divided into three matrix of dimensions 3 × 3. Hence, an explicit representation in this notation of a general third rank tensor $d_{ijk}$ (*i*, *j*, *k* = 1, 2, 3) is given in Eq. (3):

$$\vec{\vec{d}} = \begin{pmatrix} \begin{pmatrix} d_{111} & d_{121} & d_{131} \\ d_{112} & d_{122} & d_{132} \\ d_{113} & d_{123} & d_{133} \end{pmatrix} \\ ------ \\ \begin{pmatrix} d_{211} & d_{221} & d_{231} \\ d_{212} & d_{222} & d_{232} \\ d_{213} & d_{223} & d_{233} \end{pmatrix} \\ ------ \\ \begin{pmatrix} d_{311} & d_{321} & d_{331} \\ d_{312} & d_{322} & d_{332} \\ d_{313} & d_{323} & d_{333} \end{pmatrix} \end{pmatrix} \quad (3)$$

where the first index "$i$" in the tensor ($d_{ijk}$) corresponds to the rows in the main matrix (the external one). It follows then that all the elements in the first row of the inner 3 × 3 matrix have $d_{1jk}$ indices, whereas for the second row it will be $d_{2jk}$ and so on. In the same way the indices "$j$" and "$k$" will correspond to the usual way of labeling a 3 × 3 matrix, where the indices "$j$" and "$k$" are respectively the rows and columns in the inner 3 × 3 matrix.

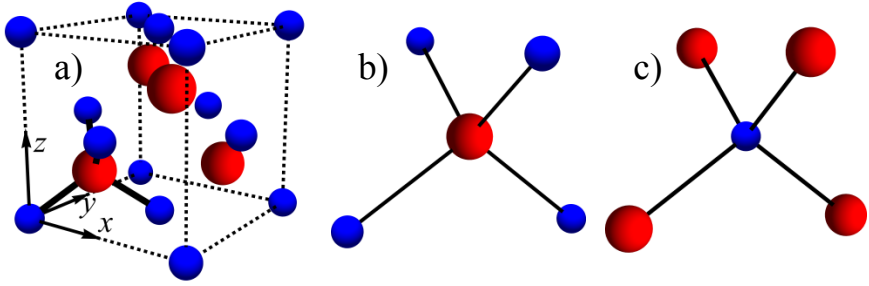

Fig. 1. a) *GaAs* conventional cell, *Ga* (red) and *As* (blue). b) Tetrahedral element, *Ga* in the center. c) Tetrahedral element, *As* in the center

*SBHM of GaAs bulk*

We start our analysis with GaAs. The conventional cell for this crystal is given in Fig. 1. Let us exemplify how formula Eq. (2) works with *GaAs* bulk crystal. The *GaAs* conventional cell is shown in Fig. 1. It is well known that *GaAs* has a conventional zincblende cell and is not centrosymmetric (See Fig. 1 a). This crystal can be imagined as generated by small tetragonal elements (Fig. 1 b). Therefore SHG will arise also from dipole contributions inside the bulk. Because SBHM is a polarizability (dipole radiation) model in the classical sense, it can be used to investigate bulk dipole contribution in such material.

If a gallium atom is chosen in the center of the pyramid, the result is going to be a susceptibility tensor $\chi_{ijk}^{Ga}$. However, it is also possible to choose As as the central position of the pyramid (Fig. 1 c). In that case the tensor will be $\chi_{ijk}^{As}$, with exactly the same tensor components. The difference between these tensors is that they have different hyperpolarizabilities $\alpha_{2Ga}$ and $\alpha_{2As}$. The bond vectors for Ga in a fixed system of coordinates (with the origin in the center of the pyramid), can be defined as follows:

$$b_1 = \begin{pmatrix} -\frac{1}{\sqrt{2}}\sin\frac{\beta}{2} \\ -\frac{1}{\sqrt{2}}\sin\frac{\beta}{2} \\ -\cos\frac{\beta}{2} \end{pmatrix} \quad b_2 = \begin{pmatrix} \frac{1}{\sqrt{2}}\sin\frac{\beta}{2} \\ \frac{1}{\sqrt{2}}\sin\frac{\beta}{2} \\ \cos\frac{\beta}{2} \end{pmatrix} \quad b_3 = \begin{pmatrix} -\frac{1}{\sqrt{2}}\sin\frac{\beta}{2} \\ \frac{1}{\sqrt{2}}\sin\frac{\beta}{2} \\ \cos\frac{\beta}{2} \end{pmatrix} \quad b_4 = \begin{pmatrix} \frac{1}{\sqrt{2}}\sin\frac{\beta}{2} \\ -\frac{1}{\sqrt{2}}\sin\frac{\beta}{2} \\ \cos\frac{\beta}{2} \end{pmatrix}, \quad (4)$$

where $\beta$ is the angle between each bond. This implies that the As bonds will have the following orientation:

$$b_5 = -b_1 \quad b_6 = -b_2 \quad b_7 = -b_3 \quad b_8 = -b_4, \quad (5)$$

and the susceptibility third rank tensor is now summed over eight bonds, with $\alpha_{2Ga}$ for the first four bonds and $\alpha_{2As}$ for the remaining four. Hence:

$$\tilde{\chi}^{(2)} = \frac{1}{V}\sum_{j=1}^{8}\alpha_{2j}\left(\mathbf{R}^{(z)}(\phi)\cdot\hat{b}_j\right)\otimes\left(\mathbf{R}^{(z)}(\phi)\cdot\hat{b}_j\right)\otimes\left(\mathbf{R}^{(z)}(\phi)\cdot\hat{b}_j\right) \quad (6)$$

and the explicit form of the tensor is

$$\begin{pmatrix} \begin{pmatrix} 0 & 0 & \alpha_{GaAs}S\sin 2\phi \\ 0 & 0 & -\alpha_{GaAs}S\cos 2\phi \\ \alpha_{GaAs}S\sin 2\phi & -\alpha_{GaAs}S\cos 2\phi & 0 \end{pmatrix} \\ \begin{pmatrix} 0 & 0 & -\alpha_{GaAs}S\cos 2\phi \\ 0 & 0 & -\alpha_{GaAs}S\sin 2\phi \\ -\alpha_{GaAs}S\cos 2\phi & -\alpha_{GaAs}S\sin 2\phi & 0 \end{pmatrix} \\ \begin{pmatrix} \alpha_{GaAs}S\sin 2\phi & -\alpha_{GaAs}S\cos 2\phi & 0 \\ -\alpha_{GaAs}S\cos 2\phi & -\alpha_{GaAs}S\sin 2\phi & 0 \\ 0 & 0 & 0 \end{pmatrix} \end{pmatrix}. \quad (7)$$

where we are defining $\alpha_{GaAs} = \alpha_{Ga} - \alpha_{As}$ and $S = \sin(\beta/2)\sin\beta$.

Therefore, the resulting tensor for the full cell will be $\chi_{ijk} = \chi_{ijk}^{Ga} - \chi_{ijk}^{As}$. This is the quantity that is measured by experiment: the effective tensor resulting of taking the difference between the two pyramidal elements. Moreover, in the case of bulk silicon this difference will clearly be zero and it is consequently a centrosymmetric cell.

Please note that the tensor shown in Eq. (7) is in a rotating frame, this means that there is a dependence in the angle $\phi$. Thus, the tensor associated with a particular crystal looks very different depending on the selection of the coordinate system chosen at the beginning. To illustrate this, the angle $\phi = 0$ is evaluated and yields

$$\left( \begin{pmatrix} 0 & 0 & 0 \\ 0 & 0 & -\alpha_{GaAs}S \\ 0 & -\alpha_{GaAs}S & 0 \end{pmatrix}, \begin{pmatrix} 0 & 0 & -\alpha_{GaAs}S \\ 0 & 0 & 0 \\ -\alpha_{GaAs}S & 0 & 0 \end{pmatrix}, \begin{pmatrix} 0 & -\alpha_{GaAs}S & 0 \\ -\alpha_{GaAs}S & 0 & 0 \\ 0 & 0 & 0 \end{pmatrix} \right) \quad (8)$$

whereas for $\phi = \pi/4$ it will give:

$$\left( \begin{pmatrix} 0 & 0 & \alpha_{GaAs}S \\ 0 & 0 & 0 \\ \alpha_{GaAs}S & 0 & 0 \end{pmatrix}, \begin{pmatrix} 0 & 0 & 0 \\ 0 & 0 & -\alpha_{GaAs}S \\ 0 & -\alpha_{GaAs}S & 0 \end{pmatrix}, \begin{pmatrix} \alpha_{GaAs}S & 0 & 0 \\ 0 & -\alpha_{GaAs}S & 0 \\ 0 & 0 & 0 \end{pmatrix} \right). \quad (9)$$

and it is easy to see from Eq. (7) that for $\phi = \pi/8$, all the components different from zero in the tensor are going to remain in the same position.

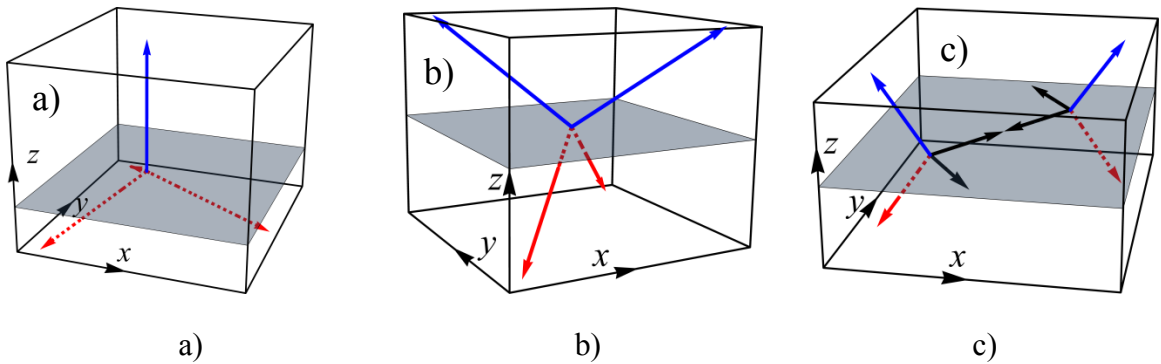

Fig. 2. Bond orientation according to the surface for Si: a) (111) b) (001) and c) (011)

### SBHM of Si surfaces

Now, following the same procedure described above, we are going to calculate the susceptibility tensor for Si(111), Si(001) and Si(011) surfaces. Fig. 2 shows the different configurations for the bonds according to the particular plane that defines the surface where we take as example the bond vectors for the Si(111) surface in Eq. (10). The details of the definitions for the bond vectors for others surfaces can be found in Ref. 9 and 11.

$$b_1 = \begin{pmatrix} 0 \\ 0 \\ 1 \end{pmatrix} \quad b_2 = \begin{pmatrix} \sin\beta \\ 0 \\ \cos\beta \end{pmatrix} \quad b_3 = \begin{pmatrix} -\frac{1}{2}\sin\beta \\ \frac{\sqrt{3}}{2}\sin\beta \\ \cos\beta \end{pmatrix} \quad b_4 = \begin{pmatrix} -\frac{1}{2}\sin\beta \\ -\frac{\sqrt{3}}{2}\sin\beta \\ \cos\beta \end{pmatrix} \quad (10)$$

Here, $\beta$ is as mentioned previously the bond angle for Si which is 109.47°. There are two different hyperpolarizabilities for the bonds pointing outside the surface $\alpha_u$ (in the $z$ direction) and for the ones pointing inside to the bulk $\alpha_l$. In the case of a Si(011) surface, the bonds in plane have a different value but cancel each other. Performing SBHM calculation, the susceptibility third rank tensor for Si(111) surface is:

$$\begin{pmatrix} \begin{pmatrix} \frac{3\alpha_l}{4}\sin^3\beta\cos3\phi & \frac{3\alpha_l}{4}\sin^3\beta\sin3\phi & \frac{3\alpha_l}{2}\sin^2\beta\cos\beta \\ \frac{3\alpha_l}{4}\sin^3\beta\sin3\phi & -\frac{3\alpha_l}{4}\sin^3\beta\cos3\phi & 0 \\ \frac{3\alpha_l}{2}\sin^2\beta\cos\beta & 0 & 0 \end{pmatrix} \\ \begin{pmatrix} \frac{3\alpha_l}{4}\sin^3\beta\sin3\phi & -\frac{3\alpha_l}{4}\sin^3\beta\cos3\phi & 0 \\ -\frac{3\alpha_l}{4}\sin^3\beta\cos3\phi & -\frac{3\alpha_l}{4}\sin^3\beta\sin3\phi & \frac{3\alpha_l}{2}\sin^2\beta\cos\beta \\ 0 & \frac{3\alpha_l}{2}\sin^2\beta\cos\beta & 0 \end{pmatrix}, \\ \begin{pmatrix} \frac{3\alpha_l}{2}\sin^2\beta\cos\beta & 0 & 0 \\ 0 & \frac{3\alpha_l}{2}\sin^2\beta\cos\beta & 0 \\ 0 & 0 & \alpha_u + 3\alpha_l\cos^3\beta \end{pmatrix} \end{pmatrix} \quad (11)$$

whereas for the Si(001) surface we have the respective tensor:

$$\left(\begin{pmatrix} 0 & 0 & S\left[\alpha_u \cos^2\phi + \alpha_l \sin^2\phi\right] \\ 0 & 0 & -\frac{3\alpha_l}{4}\sin^3\beta\cos 3\phi \\ S\left[\alpha_u \cos^2\phi + \alpha_l \sin^2\phi\right] & -\frac{3\alpha_l}{4}\sin^3\beta\cos 3\phi & 0 \end{pmatrix}\right.$$
$$\begin{pmatrix} 0 & 0 & \frac{1}{2}(\alpha_u - \alpha_l)S\sin 2\phi \\ 0 & 0 & S\left[\alpha_l \cos^2\phi + \alpha_u \sin^2\phi\right] \\ \frac{1}{2}(\alpha_u - \alpha_l)S\sin 2\phi & S\left[\alpha_l \cos^2\phi + \alpha_u \sin^2\phi\right] & 0 \end{pmatrix},$$
$$\left.\begin{pmatrix} S\left[\alpha_u \cos^2\phi + \alpha_l \sin^2\phi\right] & \frac{1}{2}(\alpha_u - \alpha_l)S\sin 2\phi & 0 \\ \frac{1}{2}(\alpha_u - \alpha_l)S\sin 2\phi & S\left[\alpha_u \cos^2\phi + \alpha_l \sin^2\phi\right] & 0 \\ 0 & 0 & 2(\alpha_u + \alpha_l)\cos^2\left(\frac{\beta}{2}\right) \end{pmatrix}\right) \quad (12)$$

where $S$ is defined as before. For the Si (011) surface the tensor is:

$$\left(\begin{pmatrix} 0 & 0 & 2\alpha_{eff}\cos^2\phi \\ 0 & 0 & \alpha_{eff}\sin 2\phi \\ 2\alpha_{eff}\cos^2\phi & \alpha_{eff}\sin 2\phi & 0 \end{pmatrix}\right.$$
$$\begin{pmatrix} 0 & 0 & \alpha_{eff}\sin 2\phi \\ 0 & 0 & 2\alpha_{eff}\sin^2\phi \\ \alpha_{eff}\sin 2\phi & 2\alpha_{eff}\sin^2\phi & 0 \end{pmatrix},$$
$$\left.\begin{pmatrix} 2\alpha_{eff}\cos^2\phi & \alpha_{eff}\sin 2\phi & 0 \\ \alpha_{eff}\sin 2\phi & 2\alpha_{eff}\sin^2\phi & 0 \\ 0 & 0 & 4\alpha_{eff} \end{pmatrix}\right) \quad (13)$$

where $\alpha_{eff} = \sqrt{2/3}(\alpha_u - \alpha_l)/3$. The latter surface needs to be handled carefully because the center of inversion is such that the conventional cell cannot be considered as a single atomic cell with four bonds but should contain two atoms with a total of eight bonds. Also, in Eq. (13) we have evaluated the angle between the bonds ($\beta = 109.47°$) to express the tensors in a more compact form.

On the other hand, since the original work by Powell *et al.*, the SBHM approach provoked criticism in the literature [12] focusing mainly on the issue that the physical foundations of SBHM were over-simplified and requires modification. As an answer to this, Aspnes and co-workers extended the original SBHM [1, 13]. However the physical model becomes more complex and direct physical interpretation is severely hidden under the underlying mathematics.

Interestingly, one aspect of the SBHM that has not been discussed before in detail is the comparison between the third rank tensor that can be obtained from group theory calculations using the symmetry of the surface and from the model itself. The next section is devoted to this topic.

### III. Group Theory and Neumann's Principle

Group Theory (GT) is not a model to describe SHG but a mathematical structure than can be applied to the symmetry operations allowed by a crystal. These symmetry operations must preserve the crystal properties, which includes the optical ones. Rotations and reflections can be represented mathematically by a matrix and for a particular crystal only rotation for specific angles are allowed for some specific existing mirror planes. The set of matrices that represents all the rotations and mirror planes for a particular crystal is a group, more specifically called a Point Group.

On the other hand, the Neumann's principle states the following [14]:

*"The symmetry elements of any physical property of a crystal must include the symmetry elements of the point group of the crystal."*

Therefore, the symmetry of the second as well as the third rank susceptibility tensors must include all of the symmetry operations contained in the point group of the crystal. This is well known and can be consulted in the classical book of Nye [14] or in books with applications of group theory [15-17]. Also, tables with the elements of the susceptibility tensor for the crystal classes are given in classical books about nonlinear optics [18, 19].

In the following discussion, we will first focus our analysis on the bulk structure of diamond like conventional cells and show that SBHM and GT will give a perfect agreement. Afterwards we discuss different groups of symmetry for the lower Si surface indices, beginning with Si(111) which has the minor number of tensorial independent parameters (four), followed by Si(001) and at the end Si(011). This is because the latter has an additional complexity as we will see below.

### *GaAs Bulk*

As we showed before, *GaAs* can be decomposed in tetragonal elements. These pyramidal elements have $T_d$ symmetry group with only one independent element in the susceptibility third rank tensor ($\chi_{123} = d_{132}$) [15]. According to Ref. 15 the tensor associated with this symmetry is:

$$\left(\begin{pmatrix} 0 & 0 & 0 \\ 0 & 0 & d_{132} \\ 0 & d_{132} & 0 \end{pmatrix}\begin{pmatrix} 0 & 0 & d_{132} \\ 0 & 0 & 0 \\ d_{132} & 0 & 0 \end{pmatrix}\begin{pmatrix} 0 & d_{132} & 0 \\ d_{132} & 0 & 0 \\ 0 & 0 & 0 \end{pmatrix}\right), \qquad (14)$$

where $d_{132}$ is just and arbitrary constant. Comparing the last expression with Eq. (8) it is clear that the tensors predicted by GT and SBHM are totally equivalent in this case and both have only one independent element. This free parameter has a physical meaning and it is nothing else but the hyperpolarizability.

*Si(111) Surface*

According to Fig. 2 a, the symmetry point group for Si(111) is $C_{3v}$. At the surface, the symmetry operations do not involve the *z*-axis because mirror planes in the *z* direction or rotations that transform in some way the *z* coordinate are forbidden due to surface orientation normal to the *z*-axis. Therefore, the third rank susceptibility tensor for this surface must include the same symmetry in accordance with the Neumann's principle.

On the other hand, if the azimuthal angle $\phi$ is fixed to some arbitrary value, the resulting susceptibility tensors in Eq. (11) will be, in the general case, different from the tensors for the symmetry groups given in the literature (Refs. 14-19). This seems to be the case why in crystals that produce SHG, the authors are not conclusive about the point group symmetry for the same low index planes of silicon surfaces discussed here [20]. Another, reason for this discrepancy is that some authors considered not only the atoms in the surface but also the second layer below the surface, getting a different point group instead [21].

The explanation for this mismatch, between one particular surface and the tensor associated to it, is -from our point of view- due to the reason that the tensors listed in the literature are usually calculated by taking the mirror planes containing two of the cartesian axes and also the rotation axis could be different for each point group with the final tensor having different symmetric distribution of its elements. In contrast, the system of reference in the SBHM is usually chosen in such a way that the unit vectors describing the bonds have *as few components as possible* to make calculations simpler. For instance, $C_{3v}$ point group has four independent elements and a total of eleven non-zero tensor elements for the conventional unit cell ($\chi_{112} = \chi_{121} = -2d_{222}$, $\chi_{113} = \chi_{131} = \chi_{223} = \chi_{232} = d_{131}$,

$\chi_{211} = -\chi_{222} = -d_{222}$, $\chi_{311} = \chi_{322} = d_{311}$ and $\chi_{333} = d_{333}$) [15], whereas Ref. 14 gives, in some cases, two possible configurations for the same point group.

In this work, we follow Ref. 15. Therefore, for the Si(111) surface tensor elements we fix the angle $\phi = \pi/2$ in Eq. (11) and obtain $C_{3v}$ symmetry tensor reported in Ref. 15, which is reproduced below:

$$\left( \begin{pmatrix} 0 & -2d_{222} & d_{131} \\ -2d_{222} & 0 & 0 \\ d_{131} & 0 & 0 \end{pmatrix} \begin{pmatrix} -d_{222} & 0 & 0 \\ 0 & d_{222} & d_{131} \\ 0 & d_{131} & 0 \end{pmatrix} \begin{pmatrix} d_{311} & 0 & 0 \\ 0 & d_{311} & 0 \\ 0 & 0 & d_{333} \end{pmatrix} \right). \tag{15}$$

However, there is a small difference that is in the form of a multiplication constant of 2 in some of the tensor elements for the $C_{3v}$ symmetry from our work and the one reported in Ref. 15 and it does not appear formally when the symmetry operations of this particular group are applied to the general tensor in order to simplify it (Electronic mail correspondence has been performed with the author to correctly address this possible typographical error) [22]. Here the coefficient $\chi_{112}$ satisfies $\chi_{112} = \chi_{121} = -d_{222}$. Whereas, for the Si(001) and Si(011) surfaces, a $C_{2v}$ tensor [15] is obtained with an angle $\phi = 0$:

$$\left( \begin{pmatrix} 0 & 0 & d_{131} \\ 0 & 0 & 0 \\ d_{131} & 0 & 0 \end{pmatrix} \begin{pmatrix} 0 & 0 & 0 \\ 0 & 0 & d_{232} \\ 0 & d_{232} & 0 \end{pmatrix} \begin{pmatrix} d_{311} & 0 & 0 \\ 0 & d_{322} & 0 \\ 0 & 0 & d_{333} \end{pmatrix} \right). \tag{16}$$

In order to compare the tensors for a fixed coordinate system of reference with the ones generated by SBHM, these tensors have to be rotated by an arbitrary angle $\phi$. This can be done using the following relation:

$$\chi_{ijk}(\phi) = R_{il}(\phi) R_{jm}(\phi) R_{kn}(\phi) \chi_{lmn}. \tag{17}$$

The resulting tensors for $C_{3v}$ and $C_{2v}$ point groups are

$$\begin{pmatrix} \begin{pmatrix} d_{111}\cos 3\phi & d_{111}\sin 3\phi & d_{131} \\ d_{111}\sin 3\phi & -d_{111}\cos 3\phi & 0 \\ d_{131} & 0 & 0 \end{pmatrix} \\ \begin{pmatrix} d_{111}\sin 3\phi & -d_{111}\cos 3\phi & 0 \\ -d_{111}\cos 3\phi & -d_{111}\sin 3\phi & d_{131} \\ 0 & d_{131} & 0 \end{pmatrix} \\ \begin{pmatrix} d_{311} & 0 & 0 \\ 0 & d_{311} & 0 \\ 0 & 0 & d_{333} \end{pmatrix} \end{pmatrix} \tag{18}$$

and

$$\begin{pmatrix} \begin{pmatrix} d_{131}\cos^2\phi + d_{232}\sin^2\phi & 0 & 0 \\ 0 & 0 & \frac{1}{2}(d_{131}-d_{232})\sin 2\phi \\ 0 & \frac{1}{2}(d_{131}-d_{232})\sin 2\phi & 0 \end{pmatrix} \\ \begin{pmatrix} 0 & 0 & \frac{1}{2}(d_{131}-d_{232})\sin 2\phi \\ 0 & 0 & d_{131}\sin^2\phi + d_{232}\cos^2\phi \\ \frac{1}{2}(d_{131}-d_{232})\sin 2\phi & d_{131}\sin^2\phi + d_{232}\cos^2\phi & 0 \end{pmatrix} \\ \begin{pmatrix} d_{311}\cos^2\phi + d_{322}\sin^2\phi & \frac{1}{2}(d_{311}-d_{322})\sin 2\phi & 0 \\ \frac{1}{2}(d_{311}-d_{322})\sin 2\phi & d_{311}\sin^2\phi + d_{322}\cos^2\phi & 0 \\ 0 & 0 & d_{333} \end{pmatrix} \end{pmatrix}, \tag{19}$$

respectively. For the $C_{3v}$ symmetry, the tensor is rotated such that the system of reference has a mirror plane perpendicular to y-axis. In order to compare the tensor with Ref. 15, the rotation angle should be evaluated at $\phi = \pi/2$ with $d_{111} \to d_{222}$ whereas for the $C_{2v}$ symmetry we simply set $\phi = 0$. Comparing Eq. (11) with Eq. (18), we can solve a system of equations for the unknown tensorial coefficients $d_{ijk}$ in terms of the physical constants from SBHM. This system of equations has more unknowns than equations so its consistency was checked. The solutions for the $C_{3v}$ case yield:

$$d_{111} \to \frac{3}{4}\alpha_l \sin^3 \beta, \quad d_{131} = d_{311} \to \frac{3}{2}\alpha_l \cos \beta \sin^2 \beta, \quad \text{and} \quad d_{333} \to \alpha_u + 3\alpha_l \cos^3 \beta. \quad (20)$$

As was mentioned before, the tensor predicted by GT has four independent elements, whereas SBHM only requires two. To go further, here we are going to introduce a new condition: Kleinman symmetry. It is defined for the case when anomalous dispersion can be ignored, then the susceptibility remains unchanged when the frequencies of the two input signals and the resulting beam are permuted. This has as a consequence that the indices in the susceptibility tensor must be invariant to all permutations [15, 18]. Mathematically this can be stated as

$$d_{ijk} = d_{jki} = d_{kij} = d_{ikj} = d_{jik} = d_{kji}. \quad (21)$$

Therefore, if Kleinman symmetry holds (which is an inherent condition in SBHM [12]), then the original four independent parameters are reduced to three but in Eq. (20) we can observe that SBHM relates two of these three independent elements by:

$$\frac{d_{111}}{d_{131}} = \frac{1}{2}\tan \beta \quad (22)$$

this means that the ratio between these components of the tensor is a constant and only depends of particular crystal used because $\beta$ is the bond angle, with a fixed value for each material. The validity of this particular result should be tested experimentally or using *ab initio* calculations.

*Si(001) Surface*

The Si(001) surface has $C_{2v}$ symmetry as can be deduced from Fig. 2 b. The tensor associated to this point group has five independent elements and a total of seven non-zero elements ($d_{113} = d_{131}, d_{223} = d_{232}, d_{311}, d_{322}$ and $d_{333}$) according to Ref. 15 and Eq. (16), whereas Ref. 14 gives in some cases two possible configurations for the same point group (but keeping the same number of independent parameters).

By fixing the angle $\phi = 0$ in Eq. (12) and Eq. (19) we obtain the tensor elements in Ref. 15. In general, we can do the same for the Si(111) surface and get the relations between the tensor elements from GT and SBHM:

$$d_{131} = d_{311} \to \alpha_u S, \quad d_{232} = d_{322} \to \alpha_l S \quad \text{and} \quad d_{333} \to 2(\alpha_u + \alpha_l)\cos^2\left(\frac{\beta}{2}\right), \quad (23)$$

where $S = \sin(\beta/2)\sin \beta$. As in the previous case, the five independent parameters can be reduced first to three by applying the Kleinman condition and then the remaining one, required by SBHM, is seen in the third equation above. If the third relation is correct has to be checked experimentally or by *ab initio* theories.

*Si(011) Surface*

The Si(011) surface conventional cell consist of two atoms and the center of inversion is located in the middle between two connecting bonds which has $C_2$ symmetry, as can be seen in Fig. 2 c. This surface has been studied using SHG by several groups. In the early work of Driscoll and Guidotti [20], the crystal class of the surface is not certain, and they argued that it could be $C_2$, $C_{2v}$, or even $C_{3v}$. The difficulty to determine the symmetry of this surface can arise due to experimental issues where terraces are formed on the surfaces thus exposing the layer below. Therefore, the real surface will be a combination of both the uppermost layer and the layer below resulting in an average symmetry of $C_{2v}$ [20, 21, 23]. In this case, SBHM agrees with all the works cited above but disagrees with Group Theory. It is clear that mathematically the (perfect) surface should have only $C_2$ symmetry but the conventional cell can be replicated itself trough two different mirror planes plus one translation by a fraction of the lattice constant (there are two possible glide planes) hence forming a nonsymmorphic space group [15]. Fig. 3 illustrated the situation describe above. A top view of the Si(011) surface, white dots are the upper layer and black dots are the layer below, whereas grey dots are in the mirror position of the black ones and the dashed dots in the mirror position of the white ones. One possible explanation, about the good agreement between the experimental results and the predictions of the SBHM, is that it is *not possible in the experiment to discriminate this additional translation*, thus resulting in a $C_{2v}$ symmetry.

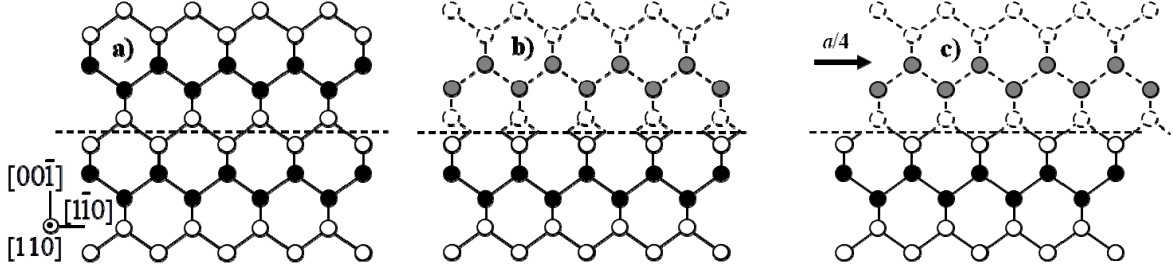

Fig. 3. Top view of Si(011) surface, the dashed line is a glide plane. a) The surface, b) the mirror plane and its image and c) a translation by a quart of the lattice constant generates the original crystal again.

In addition, optical wavelengths are very large compared to the distance between atoms. For this reason SBHM analyzes the bond vectors defined from one origin point. Such a definition of bonds will produce a $C_{2v}$ symmetry. Hence, we have to compare SBHM tensor Eq. (13) with $C_{2v}$ tensor given in Eq. (19) instead of the real symmetry of the Si(011) surface. Applying the same procedure that before produces a surprising result. Due to the linear independence of the sine and cosine functions the only possible solution in the system of equations requires

$$d_{232} = d_{322} = 0, \qquad (24)$$

which immediately sets

$$d_{311} = d_{131} = \frac{1}{2}d_{333} = 2\alpha_{eff} \qquad (25)$$

where $\alpha_{eff} = \sqrt{2/3}(\alpha_u - \alpha_l)/3$, as before. Again, we can see that from five independent parameters, Kleinman symmetry reduces them to three and the relations establish by SBHM leads to only one. Furthermore, The ratio between the nonzero components of the tensor is always a constant:

$$\frac{d_{333}}{d_{311}} = 2. \qquad (26)$$

Also, this result is very interesting, because according to SBHM, it is only possible to establish an equivalence between both tensors if the elements $d_{232}$ and $d_{322}$ are zero. In other words, when Kleinman symmetry is assumed this components of the tensor go immediately to zero. In order to verify the validity of SBHM, experiments should be performed out to check whether these elements of the susceptibility tensor are indeed zero.

One point in favor of SBHM is that, in contrast with previous approaches, it does not require any *a priori* knowledge of the surface symmetry group to make simplifications in the tensor, before contracting it with the electric fields or to expand the intensity in a Fourier series. The latter requires fitting the data with tensorial coefficients which are a function of the same azimuthal angle, giving a result that is very complicated with many parameters to fit [21, 24]. SBHM gives the right group symmetry because the geometry of the surface is implicitly stated in the vector definitions of the bonds and the direct product in Eq. (2) automatically generates a tensor belonging to the point group of that crystal.

Differences, which depend on representation, may also occur however. The tensor component that is predicted by GT- but not by SBHM, is the $d_{232}$ coefficient which according to SBHM should be zero. As we mentioned before, the validity of the SBHM prediction should be tested experimentally. It is clear that for determining this coefficient experimentally, it is not enough to use only *s*-polarization or *p*-polarization but a general state of polarization is needed.

A previous work [21] uses GT to simplify the susceptibility tensors before contracting them with the electric field and in this way, they have less parameters to fit. However, we disagree with the symmetry that they mention for the Si(001) face, which is claimed to be $C_{4v}$ on the basis that it is only seeing the atoms at the surface as hard spheres without considering the electronic charge distribution generated by the driving electric field which in this case reduce the symmetry group to $C_{2v}$. In the case of the surface Si(111) we argue that there is a typographical error, our calculation shows that $\chi_{233}$ should be zero whereas they have $\chi_{233} = \partial_{15}$. This term seems to be mixed up with $\chi_{223}$ and in this way our tensors fully agree with the ones of Ref. [21]. Once again, the shift between sine and cosine functions only implies a different choice of the coordinate system. Finally, for the Si(011) surface, which also have $C_{2v}$ symmetry, SBMH results are just a rescaling of their free parameters and instead of needing four free parameters only two are required.

## IV. Conclusions

We have analyzed SBHM comparing with Group Theory techniques, showing that even if the physical picture of this model could be over-simplified, the mathematical statements are exactly the same that in previous models. For bulk crystals, both descriptions, by either GT or SBHM are fully equivalent, delivering just one free parameter, which has to be measured experimentally or found by *ab initio* theory. Moreover, for surfaces, the symmetry point group that belongs to the surface is automatically included in the susceptibility third rank tensor when the bonds are defined. If Kleinman symmetry is assumed, the four or five free parameters from GT immediately reduced to three. SBHM imposes a more restrictive condition in the tensors reducing the number of free parameters to two. To validate these results, experiments are needed to check if SBHM is fulfilled.

**Acknowledgement**

A. A. M. would like to thank the Mexican National Council for Science and Technology (CONACyT) for financial support during this research (under contract 186852) and to Johannes Kepler University of Linz. H.H. gratefully acknowledge funding from the Austrian Ministry for Research and Education Technology Grant Southeast Asia.